\documentclass[preprint,12pt]{elsarticle} 
\usepackage{amssymb}
\usepackage{wasysym} 
\usepackage{relsize} 
\journal{Astroparticle Physics}
\def\lta{\,\raise 0.3 ex\hbox{$ < $}\kern -0.75 em
 \lower 0.7 ex\hbox{$\sim$}\,}
\def\gta{\,\raise 0.3 ex\hbox{$ > $}\kern -0.75 em
 \lower 0.7 ex\hbox{$\sim$}\,}
\newcommand{\be}{\begin{equation}}
\newcommand{\ee}{\end{equation}}
\newcommand{\rmag}{r_{\scriptstyle X}}
\newcommand{\crflux}{\Phi_{\scriptstyle CR}}
\newcommand{\ndot}{{\dot{\cal N}}_{\scriptstyle CR}}
\newcommand{\texp}{(\Delta{t})} 
\newcommand{\xoxy}{X_{\rm O16}}
\newcommand{\xcar}{X_{\rm C12}}

\begin{document}

\begin{frontmatter}

\title{{\bf Possible Solution to the Triple Alpha Fine-Tuning Problem: 
Spallation Reactions during Planet Formation} }

\author[1,2]{Fred C Adams}  
\address[1]{Leinweber Center for Theoretical Physics, 
Physics Department, University of Michigan, Ann Arbor, MI 48109} 
\address[2]{Astronomy Department, University of Michigan, Ann Arbor, MI 48109} 

\vspace{10pt}

\begin{abstract}
Carbon is produced during the helium burning phase of sufficiently
massive stars through the triple alpha process. The $0^+$ energy level
of the carbon nucleus allows for resonant nuclear reactions, which act
to greatly increase the carbon yields compared to the non-resonant
case.  Many authors have argued that small changes to the energy level
of this resonance would lead to a significantly lower carbon abundance
in the universe, and this sensitivity is often considered an example
of fine-tuning. By considering spallation reactions occuring during
the process of planet formation, this paper presents a partial
solution to this triple alpha fine-tuning problem.  Young stellar
objects generate substantial luminosities of particle radiation
(cosmic rays) that can drive nuclear reactions through spallation.  
If the standard triple alpha process is inoperative, stars tend to
synthesize oxygen (and other alpha elements) rather than carbon.
Cosmic rays can interact with oxygen nuclei to produce carbon while
planets are forming. The resulting carbon abundances are significant,
but much smaller than those observed in our universe. However, for a
range of conditions --- as delineated herein --- spallation reactions
can result in carbon-to-oxygen ratios roughly comparable to those
found on Earth and thereby obviate the triple alpha fine-tuning
problem.
\end{abstract}


\end{frontmatter} 

\newpage 

\section{Introduction} 
\label{sec:intro} 

The laws of physics, as realized in our universe, are characterized by
a collection of fundamental constants that set the strength of the
forces and the masses of particles; additional cosmological quantities
specify the inventory of the universe \cite{reessix,barrow2002}. All
of these parameters have the proper values to allow for a long-lived
universe that can develop a range of cosmic structures, including
galaxies, stars, planets, complex nuclei, and even life. A great deal
of previous work has considered the possibility that the fundamental
constants could have different values, and has explored the degree to
which they can vary and still allow the universe to be habitable
\cite{carrrees,bartip,hogan,tegmark,schellekens,donoghue,adams2019}. 
Within this enterprise, an important issue is the production of
carbon, an essential ingredient for the development of life, at least
in terrestrial forms. Variations in the fundamental constants can
lead to corresponding variations in the resonance levels of the carbon
nucleus, and can significantly change the carbon production rate
during stellar nucleosynthesis. The resulting carbon abundance depends
rather sensitively on the carbon resonance structure and hence on the
underlying fundamental constants. This sensitivity is often cited as
an example of fine-tuning, where small changes to the fundamental
constants could lead to large changes in carbon production and lead 
to a lifeless universe. This paper explores an alternate channel of
carbon production through spallation reactions that take place during
the planet formation process and thus offers a way to avoid the triple
alpha fine-tuning problem.
 
Carbon is synthesized in stars during their helium burning phases and
occurs through the triple alpha process \cite{clayton,kippenhahn}.
The expected reaction, where two $^4$He nuclei are forged into a
$^8$Be nucleus, is compromised because the $^8$Be nucleus is unstable,
with a half-life of only $t_{1/2}\sim10^{-16}$ sec. Even with this
short half-life, however, the stellar core builds up a small transient
population of $^8$Be, which can then interact with additional alpha
particles to produce carbon \cite{salpeter1952}. The latter reaction
occurs rapidly enough to account for the carbon inventory of our
univese, but requires a resonance in the carbon nucleus
\cite{hoyle1954}, where the resonance energy is 7.644 MeV
\cite{dunbar1953}, and corresponds to a $0^+$ nuclear state of the
carbon nucleus.  If the energy level of the resonance were higher
(lower), then the reaction rate could be slower (faster), resulting 
in different carbon yields. 

Previous work has explored the sensitivity of carbon production to the
specific value of the carbon resonance energy level. These studies
show that the energy level of the resonance is sensitive to the values
of the fundamental constants \cite{epelbaum2013,epelbaum2011} and that
the carbon yields in stars are sensitive to the value of the resonance
\cite{livio1989,oberhummer2000,huang}. If the resonance energy is
higher, with $(\Delta{E}_R)>0$, then carbon yields from massive stars
decrease.  In contrast, lower resonance levels with $(\Delta{E}_R)<0$
lead to increased carbon production. The range of resonance energy
that allows for net carbon production is approximately --300 keV
$\le(\Delta{E}_R)\le$ 500 keV, or a total span of $\sim800$ keV
\cite{huang}. This range should be compared to the spacing of energy
levels in the carbon nucleus, about 3 MeV, which implies roughly a one
in four chance of the resonance energy falling in the proper range to
allow for carbon production.  As another point of comparison, the
$^8$Be nucleus only fails to be stable by 92 keV.  As a result, the
changes to the energy levels of nuclei required to compromise the
triple alpha process are almost an order of magnitude larger than the
changes needed to make $^8$Be stable. The latter possibility allows
for carbon production without any need for the triple alpha process
\cite{ag2017}.

The above discussion applies to the {\it net} amount of carbon
produced in stars. For the parameter choices that lead to low carbon
yields, the stars still process alpha particles into carbon.  The
issue is that the nuclear burning temperature required for carbon
production becomes large enough that newly synthesized carbon is
promptly processed into oxygen (e.g., through the reaction $^{12}$C +
$^4$He $\to$ $^{16}$O), so that relatively little carbon remains. For
even more extreme parameter choices, much of the oxygen can be
processed into larger alpha elements such as neon, magnesium, and
silicon. As a result, when stars fail to produce substantial amounts
of carbon, they leave behind oxygen and other alpha elements instead.
These nuclei can be broken down to produce carbon through spallation
reactions.

Note that this paper implicitly considers variations in the energy
levels of carbon of order 100 to 500 keV. Although the binding
energies of carbon, oxygen, and other alpha elements are expected to
also change by roughly comparable increments, such changes are small
in relative terms. Recall that the binding energy of carbon is about
92 MeV, and that for oxygen is about 124 MeV \cite{clayton}. As a
result, changes to the binding energies are small, even though large
changes to the carbon production rates can be realized.

By considering spallation reactions that convert oxygen into carbon
during the process of planet formation, this paper constructs an
alternate mechanism to circumvent the triple alpha fine-tuning
problem. Planets form in the circumstellar disks associated with young
stellar objects during the first $\sim1-10$ Myr of evolution. During
this time span, the stellar hosts produce copious amounts of particle
radiation --- cosmic rays --- which have the potential to change the
nuclear composition of the planet-forming material through
spallation. In the model explored here, strong magnetic fields couple
to the disk at radii $r\sim\rmag\sim0.1$ AU, where reconnection events
lead to particle acceleration and large fluxes of cosmic rays (Section
\ref{sec:radiate}). This particle radiation can drive spallation
reactions that convert some fraction of the oxygen nuclei into carbon
(Section \ref{sec:o2c}). Because the cosmic rays originate near the
inner edge of the disk, planets forming with shorter orbital periods
tend to have higher carbon abundances.  For small stars, planets in
the habitable zone can have carbon to oxygen ratios comparable to that
of Earth (Section \ref{sec:hzone}).  As summarized in Section
\ref{sec:conclude}, this scenario allows for the development of
potentially habitable planets, even in universes where the triple
alpha process cannot produce high enough carbon yields through stellar
nucleosynthesis.

\section{Production of Particle Radiation during Early Stellar Evolution} 
\label{sec:radiate} 

During their early formative phases, young stars are accompanied by
circumstellar disks that provide the initial conditions for the planet
formation process. These disks are truncated near the host star due to
the strong magnetic fields generated within the star.  The magnetic
truncation radius $\rmag$ is determined by the balance between the
inward pressure due to mass accretion through the disk and the outward
pressure due to the fields, and can be written in the form  
\be
\rmag = \eta \left( {B_\ast^4 R_\ast^{12} \over 
G M_\ast {\dot M}^2} \right)^{1/7} \,,
\label{trunk} 
\ee
where ${\dot M}$ is the mass accretion rate and $B_\ast$ is the field
strength on the stellar surface.  Other parameters include the stellar
radius $R_\ast$ and the stellar mass $M_\ast$. Finally, $\eta$ is a
dimensionless parameter that is expected to be of order unity, and its
value varies with the model details \cite{ghoshlamb,blandford}.  For
typical pre-main-sequence stars, surface field strengths fall in the 
range $B_\ast\approx1-2$ kilogauss, so that truncation radii
$\rmag\sim0.1$ AU (note that all of the physical quantities appearing
in equation [\ref{trunk}] have been measured for a large number of
pre-main-sequence stars \cite{jkrull}).

In the astrophysical model under consideration \cite{shu1997,lee1998},
particle acceleration occurs due to magnetic reconnection events that
take place near the truncation radius of equation (\ref{trunk}). The
strong magnetic field lines tend to make the star co-rotate with the
inner edge of the disk (at $\rmag$), but fluctuations in the mass
accretion rate, the magnetic field strength, and other parameters
require the field lines to continually adjust through reconnection
events. This magnetic activity, in turn, leads to particle
acceleration. This process is essentially a scaled-up version of the
mechanism that produces cosmic radiation in the Sun \cite{meyer1956}.
The luminosity $L_p$ in energetic particles is expected to be
comparable to, but somewhat less than, the luminosity $L_X$ emitted in
X-rays \cite{padovani2016}.  The latter luminosity is observed to be a
substantial fraction of the photon luminosity $L_\ast$, with both 
temporal and source-to-source variations \cite{feigelson,preibisch}. 
As a result, the power contributions from young stellar objects are 
ordered according to the relation  
\be
L_p \sim 10^{-1} L_X \sim 10^{-4} L_\ast \,. 
\label{luminosity} 
\ee
Note that pre-main-sequence stars are brighter than their main
sequence counterparts, and that low mass stars evolve more slowly than
larger stars during PMS contraction. As a result, the typical stellar
luminosity is given by $L_\ast\sim1L_\odot$ during the epoch of planet
formation (stellar ages $t\lta10$ Myr).

Cosmic rays generally have a power-law energy distribution 
\cite{reames1997,reames1999}, which can be written in the form 
\be
{dN \over dE} = f(E) = {p-1 \over E_0} 
\left({E_0 \over E}\right)^p \,,
\ee
where the distribution is normalized over the range
$E_0\le{E}\le\infty$. For cosmic radiation generated in the Sun
through gradual flares, the power-law index $p\approx2.7$
\cite{vanhollebeke}. For particle radiation generated through
implusive flares, the distribution is steeper with $p\approx3.5$. For
the sake of definiteness, we use $p=3$ for the estimates of this paper
(for completeness, note that the distribution for galactic cosmic rays
has energy dependence $\sim E^{-2.7}$ at high energies, which is
roughly similar). The luminosity relations of equation
(\ref{luminosity}) correspond to cosmic rays with energies
$E\ge{E_0}=10$ MeV. As outlined below, the energy threshold
$E_{th}$ for the nuclear reactions of interest is comparable to, but
somewhat larger than, the scale $E_0$.  For a given energy spectrum,
the number of cosmic rays produced per unit time $\ndot$ with energies
$E\ge E_{th}$ is related to the luminosity $L_p$ according to
\be 
\ndot = {p-2 \over p-1} {L_p \over E_0} 
\left({E_0\over E_{th}}\right)^{p-1} = 
{L_p \over 2E_0} \left({E_0\over E_{th}}\right)^{2} \,,  
\ee 
where the final equality assumes the index $p=3$. Here, $L_p$ is 
the total luminosity in cosmic rays over the energy range $E\ge E_0$,
so that the factor $(E_0/E_{th})^2$ corrects for the inclusion of 
only those particles with energy $E\ge E_{th}\ge E_0$. 

In this scenario, the cosmic radiation is generated within an annulus
in the disk roughly centered on the truncation radius $\rmag$ from
equation (\ref{trunk}). As a result, the number flux (fluence) of cosmic 
rays that is generated with energy $E>E_{th}$ is given by 
\be 
\crflux \approx {\ndot \over \pi \rmag^2} = 
{L_p E_0 \over 2 \pi \rmag^2 E_{th}^2} \, \approx 
2 \times 10^{9}\, \left({E_0\over E_{th}}\right)^{2} 
\,{\rm cm}^{-2}\,{\rm s}^{-1}\,,
\label{numflux}  
\ee 
where the numerical estimate uses $E_0=10$ MeV, $L_p=10^{-4}L_\odot$,
and $\rmag=0.1$ AU. 

Equation (\ref{numflux}) specifies the number flux in the vicinity of
the truncation radius $r\sim\rmag$. The subsequent propagation of the
particles from their point of origin is complicated. Since the
magnetic field lines are continually wrapped up by the shear flow from
the disk, and then reconnected, the corresponding field structure is
expected to be complex and chaotic in the region where the cosmic rays
are accelerated.

In this setting, the disk scale height $h$ provides an important
length scale in the problem, where $h\sim\rmag/20$ (see, e.g.,
\cite{shu1997}). The cosmic rays propagate by moving along the field
lines, with a typical length scale $h$, and perpendicular to the field
lines, with typical length scale set by the magnetic gyroradius
$r_g=\gamma mcv/qB$. For typical energy (10 MeV -- 10 GeV) and field
strength ($B\sim1$ gauss), $r_g\ll{h}$, so that the particles are tied
to the magnetic field and travel along the highly convoluted field
lines. In the absence of attenuation, cosmic ray propagation can thus
be modeled as a random walk with step length $h$ (note that
attenuation is considered in the following section).  In order for
cosmic rays to leave the annulus where they are generated, they must
travel a total distance $d\sim\rmag=\sqrt{N}h$, where $N$ is the
required number of steps. As a result, the cosmic rays must propagate
across the disk scale height many times before escaping, i.e., ${\cal
  M}=(\rmag/h)^2\sim400$. The total distance traveled is given by
${\cal M}h=\rmag^2/h\sim20\rmag$. With such a long path length, a
large fraction of the cosmic rays will be absorbed within the annulus
where they are generated.

\section{Spallation of Oxygen to make Carbon in Circumstellar Disks} 
\label{sec:o2c} 

The current working scenario for rocky planet formation can be
summarized as follows. The raw materials for planet formation begin in
the form of dust grains. In the interstellar medium, a substantial
fraction of the metals (elements beyond helium) are locked up in these
entities.  Although they have a distribution of sizes, the grains
begin with a typical size $b_g\sim0.1-1\mu$m. Within the circumstellar
disks that form planets, the dust grains grow into larger bodies,
sometimes known as pebbles, with sizes $b_p\sim0.1-1$ cm. The pebbles
sink to the midplane of the disk, experience a streaming instability,
and thereby produce rocky planetesimals with sizes $b_R\sim1-100$ km
\cite{drazkowska} (alternately, planetesimals can be produced through
a gravitational instability \cite{goldward}). These planetesimals
subsequently accumulate into rocky terrestrial planets.

In our galaxy the original dust grains, and hence the rocks they grow
into, are generally composed of both silicates and graphite, with
admixtures of many other minerals \cite{drainedust}.  For the case
under consideration here, with little carbon produced in stars, the
grains are expected to be primarily silicates and thus contain a large
mass fraction of oxygen. For example, typical compounds found in solar
system rocks include SiO$_2$ and Fe$_2$O$_3$ (see \cite{lodders}).

If the rocky entities are exposed to a flux of cosmic radiation, then
spallation can produce carbon through a number of possible reactions,
e.g., 
\be
^{16}{\rm O} + p \longrightarrow \,\,\,
^{12}{\rm C} + 3p + 2n
\qquad {\rm and} \qquad 
^{16}{\rm O} + p \longrightarrow \,\,\,
^{12}{\rm C} + p + \alpha \,,
\ee
along with many others. In this case, oxygen-16 represents the target
nucleus.  For common reaction products --- such as carbon --- the
cross sections for such spallation reactions are generally of order
$\sigma_0=100$ mb \cite{silbertsao,soppera,dyer1981,lang1987,nero1973}. 
For spallation reactions that convert oxygen to carbon, the data show
that there is an effective threshold energy at $E=E_{th}\approx15$
MeV, where the cross section rises steeply as a function of energy,
and then varies slowly for larger energies. For the estimates of this
paper, we adopt the simple approximation where $\sigma(E)=\sigma_0$ =
100 mb = $10^{-25}$ cm$^2$ for $E\ge E_{th}$ and $\sigma(E)=0$
otherwise.

Carbon production is maximized if the target oxygen nuclei are exposed
to the full flux of cosmic rays, i.e., in the optically thin limit. In
practice, this condition holds if the rocky bodies are in the form of
pebbles ($b_p\sim1$ cm), whereas planetesimals ($b_R\sim1-100$ km) are
optically thick. On the other hand, if the bodies are too small
($b_g\sim1\,\mu$m), they remain coupled to the gas and are swept into 
the star along with the gas in the accretion flow. We thus consider 
intermediate sizes. These centimeter-sized bodies are expected to be
decoupled from the gas inside the truncation radius $r<\rmag$, where
the column density of the gas is low \cite{shu1997}.

With the above specifications, the rate $\Gamma$ at which a given
target oxygen nucleus interacts to make carbon is given by 
\be
\Gamma = \crflux \langle\sigma\rangle = 
{L_p \sigma_0 E_0 \over 2 \pi \rmag^2 E_{th}^2} \,. 
\label{thinrate} 
\ee
In the limit where the target nuclei are exposed to an unattenuated
cosmic ray flux for an exposure time $\texp$, the time evolution of 
the carbon/oxygen inventory is given by 
\be
{\rm [C/O]} = {4\over3} {\xcar \over \xoxy} 
= \exp[\Gamma \texp] - 1 \approx \Gamma\texp = 
{L_p \sigma_0 E_0 \texp \over 2\pi \rmag^2 E_{th}^2} 
\sim 0.0025 \,, 
\label{ratioevolve} 
\ee
where [C/O] is the ratio of the number of carbon to oxygen nuclei and
$X_k$ is the corresponding mass fraction of the nuclear species (note
that we also assume that carbon is the most common product of the
spallation reactions).  The final approximate equality holds in the
expected limit where only a small fraction of the oxygen is
converted. The numerical estimate assumes typical values for the
parameters, i.e., $L_p=10^{-4}L_\odot$, $\sigma_0$ = 100 mb, $E_0$ =
10 MeV, $E_{th}$ = 15 MeV, $\rmag=0.1$ AU, and $\texp$ = 1 Myr.

Note that the above estimate only includes reactions for carbon
production where oxygen nuclei are the target, but other targets are
possible.  In the scenario under consideration here, we are assuming
that carbon is rare because most of the carbon produced in stars is
immediately synthesized into oxygen. The expected oxygen abundance is
thus given approximately by the combined mass fraction of carbon and
oxygen in our solar system, or about $\sim2/3$ of the total mass
fraction of metals \cite{lodders}. The remaining $\sim1/3$ of the
metal mass fraction provide additional targets for carbon production
(as well as additional attenuation -- see below).  Since the
corresponding reaction cross sections are lower, however, the
correction for non-oxygen targets is expected to be modest. In any
case, the neglect of alternate targets renders the estimates of this
paper as lower limits.

The expected exposure time $\texp$ is subject to several constraints.
Cosmic ray production occurs due to reconnection events taking place
near the truncation radii of circumstellar disks, so the disk lifetime
limits the exposure time (at least for the most energetic phases).
Observed circumstellar disks have a distribution of lifetimes
\cite{jesus2007} that can be modeled as an exponential with half-life
$\sim3$ Myr. As a result, only $\sim10\%$ of disks live as long as 10
Myr, which thus provides an approximate upper limit on $\texp$. As
outlined below, another constraint arises from the growth of the rocky
material containing the oxygen targets. While the rocks are small,
they are exposed to the full flux of cosmic rays, but shielding occurs
once the rocks have grown larger than $b\sim1-10$ cm. The time at 
which rocky bodies cross this size threshold, typically a few Myr 
\cite{pebbles2021}, provides another constraint on the effective
exposure time. Calculations that explain the observed/inferred
abundances of short lived radioactive nuclei in our solar system
lead to similar time scale estimates \cite{leya2003}. In order for 
spallation to simultaneously produce the correct isotopic ratios for
$^{26}$Al/$^{27}$Al, $^{41}$Ca/$^{40}$Ca, $^{53}$Mn/$^{55}$Mn, and
$^{92}$Nb/$^{92}$Nb, the irradiation time must be of order 
$\sim1$ Myr (see also \cite{shu1997,lee1998}). 

Since the target nuclei of interest tend to be locked up in rocky
bodies, maximal exposure occurs only for sufficiently small entities.
When cosmic rays enter rock, they lose energy primarily through
electromagnetic interactions with electrons, rather than through
nuclear reactions. For example, cosmic rays with energy $E=10$ MeV
lose their energy after passing through a column density of only
$\Sigma_s\sim0.3$ g cm$^{-2}$ \cite{reedy,pavlov}, with larger
stopping columns for particles of higher energy. In addition to energy
loss, the propagation of cosmic rays through rock is complicated by
the emission of secondary particles, such as neutrons and electrons,
which can propagate longer distances (although their impact is small
in the present application). As a result, one must consider the full
energy distribution of cosmic rays and perform detailed simulations of
the process. For example, for the case of solar cosmic rays impinging
on the Martian surface, the isotope production rate for carbon
decreases with depth $z$ with an nearly exponential fall-off
\cite{pavlov}. Motivated by this finding we write the production 
rate in the approximate form 
\be
\gamma = \gamma_0 \exp[-z/\zeta] \,,  
\label{rockrate} 
\ee
where we expect the effective scale height $\zeta\sim1$ cm
\cite{reedy,pavlov} (and where this approximation is valid for 
solar cosmic rays). The corresponding rate coefficient $\gamma_0$ is
calculated for an incident flux of solar cosmic rays in the energy
range $E\ge1$ MeV. If we rescale the coefficient to include only
cosmic rays with energy $E\ge E_0=10$ MeV, and define a benchmark
number flux of protons $\Phi_1$ = 1 cm$^{-2}$ s$^{-1}$, then we find
the value $\gamma_0\approx0.003(\crflux/\Phi_1)$ atom cm$^{-3}$
s$^{-1}$, where $\crflux$ is the number flux of cosmic rays with
energy $E\ge10$ MeV.  Using these results, let us consider a cosmic
ray flux $\crflux$ incident on a slab of rocky material with thickness
$b$. The rate at which oxygen nuclei are converted into carbon is
given by the expression 
\be
\Gamma_R = \left( {16 m_p \gamma_0 \over \rho_R \xoxy} \right) 
{\zeta\over b} \left[ 1 - \exp(-b/\zeta) \right] \,, 
\label{thickrate} 
\ee
where $\rho_R=2.5$ g cm$^{-3}$ is the density of the rock. In the
limit of small thickness $b\ll\zeta$, the function on the right
approaches unity, and the rate becomes 
\be
\Gamma_R(b\ll\zeta) \approx (10^{-25}\,\,{\rm s}^{-1}) 
\left( {\crflux \over \Phi_1}\right) \,.
\label{thicklimit} 
\ee
This rate $\Gamma_R(b\ll\zeta)$ can be compared to the optically 
thin rate $\Gamma=\langle\sigma\rangle\crflux$ from equation
(\ref{thinrate}).  The rates are comparable in this limit, as
expected. For completeness, note that the optically thick rate from
equation (\ref{thickrate}) in the optically thin limit
(\ref{thicklimit}) is equal to the optically thin rate from equation
(\ref{thinrate}) for a specific choice of oxygen abundance (i.e., the
rate coefficient $\gamma_0$ depends on $\xoxy$). In any case, rocky
bodies with size $b\ll\zeta$ can be treated using the optically thin
expression. As a working approximation, we assume that the net effect
of rocky material agregating into larger bodies of size $b$ is to
reduce the effective rate of converting oxygen into carbon by the
factor $\sim(\zeta/b)(1-\exp[b/\zeta])$. As a result, the largest
rocks that remain optically thin have sizes less than about
$b\sim\zeta\sim1$ cm. 

For larger rocks, $b>\zeta$, only the outer layer is exposed to the
cosmic ray flux. Within the region of disk where the cosmic rays are
generated, strong but tangled magnetic fields confine the particle
radiation, which must propagate diffusively. As a result, the cosmic
radiation is expected to be nearly isotropic. For larger rocks,
assuming spherical geometry, the fraction ${\cal F}$ of their volume
that will experience spallation reactions is given by 
\be
{\cal F} \approx 3 \left({\zeta\over b}\right) - 
3 \left({\zeta\over b}\right)^2 + 
\left({\zeta\over b}\right)^3 
\approx {3\zeta\over b} \,, 
\label{fraction} 
\ee
where the final expression corresponds to the limit $b\gg\zeta$. 

The discussion thus far assumes that the target oxygen nuclei, or
rocky bodies that contain them, are exposed to the full flux of cosmic
rays. If the total amount of solid material in the reconnection region
is large enough, then the cosmic ray flux can be attenuated by
background before reaching the rocks of interest. We can address this
issue by estimating the optical depth of the background field of rocky
material. Let the total mass $M_s$ in solids be spread out over the
reconnection region, which is an annulus with radius comparable to the
truncation radius $\rmag$. The column density of solid material is
thus given by 
\be
\Sigma_s \approx {M_s \over \pi \rmag^2} \,. 
\ee 

The solid material is primarily contained in rocky bodies, which can
be considered as partially optically thick. We further consider the
rocks to have a given radius $b$.  As outlined above, the nuclei
within the rocks are exposed to only a fraction ${\cal F}(b)$ of 
the total cosmic ray flux, whereas the correspoding fraction 
$[1-{\cal F}(b)]$ is lost (primarily through electromagnetic 
interactions).\footnote{Note that the particles are still present, 
but they are slowed down below the energy threshold so that they 
are no longer useful for spallation reactions.} The optical depth 
of the reconnection region can then be written in the form 
\be
\tau_{back} \approx {M_s\over\pi\rmag^2}\,{3\over4\rho\,b} 
[1-{\cal F}(b)] \approx {200\over(b/1\,{\rm cm})} 
[1-{\cal F}(b)]\,,
\ee
where the numerical estimate uses benchmark values $M_s=1M_\oplus$ 
and $\rmag$ = 0.1 AU. In the limit of large bodies $b\gg\zeta$, the
fraction ${\cal F}\to0$, so that all of the cosmic rays striking a
given rocky body are lost.  In this case, the optical depth
$\tau_{back}$ is determined by the surface (number) density of rocky
bodies times their geometrical cross section. In the opposite limit of
small rocks with $b\ll\zeta$, the fraction ${\cal F}\to1$ but 
$1-{\cal F}={\cal O}(b/\zeta)$ so that the optical depth becomes
independent of the rock size $b$ (e.g., see \cite{asradio}).

The optical depth $\tau_{back}$ and the factor ${\cal F}$ jointly
determine the fraction $f_{cr}$ of the total cosmic ray flux seen by
the typical target nucleus.  If we assume a simple one-dimensional
geometry for cosmic ray propagation across the reconnection region,
then the nuclei within the rocks are exposed to a fraction of the
original cosmic ray flux given by
\be
f_{cr} = {1 \over \tau_{back}} 
\left[ 1 - {\rm e}^{-\tau_{back}} \right]
{\cal F}(b)\,.
\label{fsurvive} 
\ee
The first factor takes into account the loss of cosmic rays due to the
background sea of rocky bodies and the second factor represents the
fraction of nuclei within a given rock that are exposed to cosmic
rays. 

The survival fraction $f_{cr}$ depends on both the total amount of
mass $M_s$ in the reconnection region and the size $b$ of the
individual rocks.  For solid masses $M_s\ll1M_\oplus$, the optical
depth $\tau_{back}\ll1$ and the survival fraction $f_{cr}={\cal F}$.
If, in addition, the rocks are small so that $b\lta1$ cm, both
$f_{cr}$ and ${\cal F}$ are of order unity so that the nuclei are
exposed to nearly the full flux of cosmic rays. In this limit, the
carbon-to-oxygen ratio evolves with time according to equation
(\ref{ratioevolve}). In general, however, the interaction rate
$\Gamma$ (see also equation [\ref{thinrate}]) must be reduced by the
factor $f_{cr}$. For example, in the case of large total mass 
$M_s\gta1M_\oplus$, the optical depth is expected to be large (of
order $\sim100$), so that equation (\ref{fsurvive}) reduces to
$f_{cr}\approx{\cal F}(b)/\tau_{back}$.  Since the rocks
are (most likely) in the form of pebbles with $b\sim1$ cm, the
fraction ${\cal F}\sim1$, so that the reduction factor
$f_{cr}\sim1/\tau_{back}$. 


For completeness, we note that cosmic rays can also lose energy
through interactions with gas in the reconnection region. However,
such losses are already incorporated into this treatment, since the
cosmic ray luminosity $L_p$, as defined here, corresponds to those
particles that are emitted, i.e., the particles that survive
propagation over distances comparable to the inner disk size.
Moreover, the column density of gas in the region $r<\rmag$ is
expected to be small, $\sim10^{-4}$ g cm$^{-2}$, based on soft X-ray
observations of young stellar objects \cite{shu1997}.  For the
estimates of this paper, we take into account possible attenutation by
considering smaller values of the original (net) cosmic ray luminosity
$L_p$.

\begin{figure} 
\includegraphics[scale=0.60]{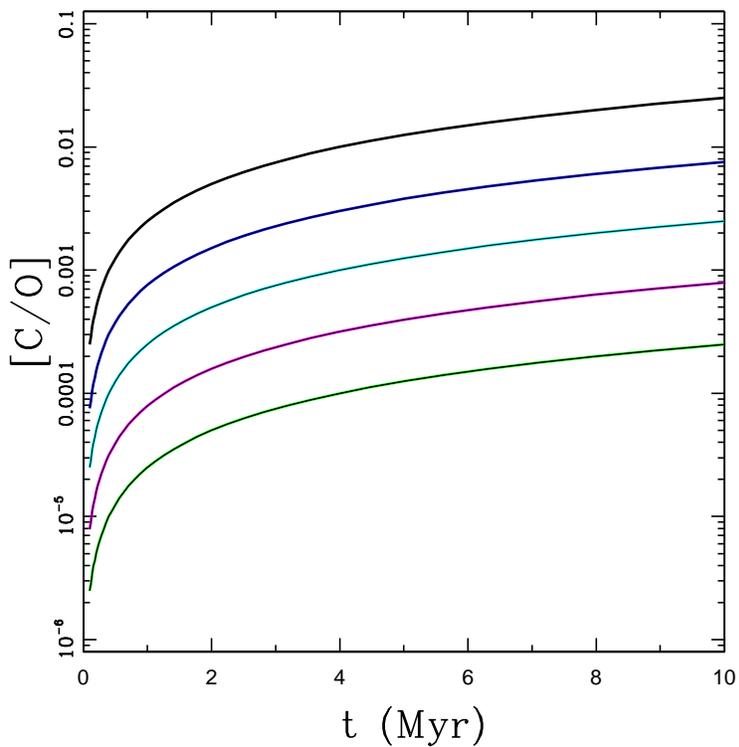}
\centering
\vskip-3.0truecm
\caption{Time evolution of the carbon to oxygen ratio [C/O] for  
varying levels of attenuation. The upper black curve shows the maximum
growth of [C/O] as a function of time for optically thin conditions
and the full expected flux of cosmic rays. The remaining curves show 
the [C/O] ratio for total optical depths given by 
$\tau_{back}\approx3.16$ (blue), 10.0 (cyan), 31.5 (magenta), and 
100 (green). For the expected case where the rocky material resides 
in pebbles ($b\sim1$ cm), these optical depths correspond to total
mass in solids $M_s/M_\oplus\approx0.0316$, 0.1, 0.316, and 1.0,
respectively. }
\label{fig:coratio} 
\end{figure} 

Figure \ref{fig:coratio} illustrates the possible time evolution of
the carbon to oxygen ratio [C/O] for various levels of attenuation, 
corresponding to various total mass in solids in the reconnection 
region. If the rocky material resides in pebbles ($b\sim1$ cm) and
experiences the full unattenuated flux of cosmic rays, then the [C/O]
ratio evolves with exposure time according to equation
(\ref{ratioevolve}), as shown by the solid black curve. The remaining
curves show the time evolution of the [C/O] ratio for cases where the
cosmic ray flux is attenuated, with the total optical depth given by
$\log_{10}\tau$ = 0.5 (blue), 1.0 (cyan), 1.5 (magenta), and 2.0
(green).  These optical depth values correspond to total masses in 
solid material of $M_s/M_\oplus=10^{-3/2}$, $10^{-1}$, $10^{-1/2}$,
and 1.0. The rocky material, which initially resides in small pebbles,
eventually transitions into planetesimals with sizes $R$ = 10 -- 100
km. At this point, the the rate of transforming oxygen into carbon
decreases abruptly by a factor of $\sim10^4-10^5$. As a result, the
[C/O] ratio effectively freezes out at the transition time, i.e., the
transition time defines the effective exposure time. 

We can summarize this scenario for carbon production as follows: The
rocky material is expected to reside in the form of pebbles with size
$b\sim1$ cm and the total mass in the reconnection region is
comparable to that of a small rocky planet (so that 
$M_s\sim0.1-1M_\oplus$). Finally, the exposure time falls in the range
$\texp=1-10$ Myr, consistent with observed disk lifetimes. Under these 
conditions, the optical depth of the background is large, 
$\tau_{back}\gg1$, so that the survival fraction 
$f_{cr}\propto \tau_{back}^{-1}\propto M_s^{-1}$. Moreover, only a
small fraction of the total number of oxygen nuclei will be processed
into carbon. In this regime of parameter space, the carbon to oxygen
ratio can be written in the form 
\be
{\rm [C/O]} \approx 0.00025 \left({\texp\over10\,{\rm Myr}}\right)
\left({M_s\over M_\oplus}\right)^{-1}\,. 
\ee

\section{Planet Formation and Habitable Zones} 
\label{sec:hzone} 

The previous sections show that cosmic rays are accelerated near the
truncation radius $\rmag$, are confined to nearby regions, and can
interact with oxygen nuclei to make carbon. This section briefly
considers the conditions necessary for the resulting carbon to be
incorporated into a potentially habitable planet.

Even under the most favorable conditions, the carbon abundance is
quite small, only a fraction of the total mass of the planet (which 
is made from the rocky raw material enriched via spallation). However,
the Earth itself also has relatively little carbon, in contrast to 
its large abundance in both the Sun and the Galaxy. Even the carbon
abundance for Earth's crust has been difficult to measure, with
estimates of the carbon to oxygen ratio falling in the range [C/O]
$\approx$ 0.0002 -- 0.01 \cite{allegre2001,marty2012,marty2020,
miller2019,crchandbook,mcdonough}. Estimates for the bulk composition
of Earth are more uncertain, but fall in a similar range. Finally, we
note that oxygen makes up $\sim30\%$ of the Earth's mass, so the mass
fraction of carbon is smaller than the carbon to oxygen ratio, say
$\xcar\sim0.0001-0.003$. Significantly, the [C/O] ratios that are
produced through spallation reactions (Figure \ref{fig:coratio}) can
be comparable to (and under extreme conditions larger than) the values
estimated for Earth. However, the largest [C/O] ratios only arise for
relatively small total mass $M_s$ in solids. To form planets with
$M_p=M_s=1M_\oplus$ and an exposure time of 10 Myr, we find [C/O]
$\approx0.0003$, which is near the low end of the range inferred for
Earth.

The most straightforward scenario is for the planet to form at the
annulus centered on $\rmag$ where the cosmic rays are generated.  
In this case, the carbon to oxygen ratio in the rocky material will
sample the values illustrated in Figure \ref{fig:coratio}, i.e., [C/O]
$\approx10^{-5}-10^{-2}$.  From this birth scenario, the resulting
planetary orbit has a semimajor axis near $a\sim\rmag\sim0.1$ AU. In
order for the planet to receive the same stellar flux as Earth, the
star must be less luminous than the sun by a factor of 100, which
corresponds to a main sequence star with mass $M_\ast=0.25M_\odot$.

\begin{figure} 
\includegraphics[scale=0.60]{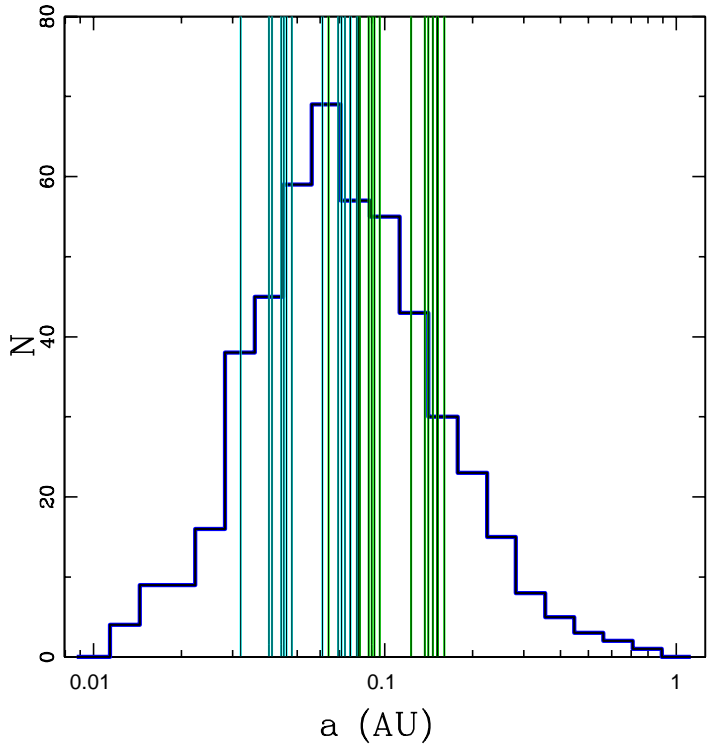}
\centering
\vskip-3.0truecm
\caption{Histogram showing semimajor axes of observed rocky planets 
(thick blue curve) and the inferred positions of the magnetic  
truncation radius (vertical lines). The planet sample includes
detected planets from the {\it Kepler} mission that have estimated
masses $M_p\le7M_\oplus$ and are found in systems containing 3 or more
planets. The green lines show the locations of the magnetic truncation
radii $\rmag$ for measured properties of pre-main-sequence stars
\cite{jkrull} according to equation (\ref{trunk}). The cyan lines show
locations corresponding to half of the inferred values ($\rmag$/2). 
The region where cosmic rays are generated, as delineated by the
collection of vertical lines, overlaps with the observed locations 
of rocky planets. } 
\label{fig:rmag} 
\end{figure} 

Figure \ref{fig:rmag} shows that rocky planets readily form near the
location of the magnetic truncation radius where cosmic rays are
generated. The blue histogram in the figure depicts the semimajor axes
for a collection of exoplanets detected by the {\it Kepler} mission
\cite{kepler}. This sample includes the planets with inferred masses
$M_p\le7M_\oplus$ that are found in systems with three or more
planets. Planets with masses below this threshold exhibit a primarily
rocky composition, whereas larger planets tend to have extended
atmospheres (although substantial variation exists). Only planets in
multi-planet systems were used, as those tend to have nearly circular
orbits and make up a uniform sample.\footnote{Keep in mind that for 
many planets the mass is inferred from measurement of the planetary
radius, i.e., the mass is not directly determined.}  The green
vertical lines show the locations of the magnetic truncation radii
$\rmag$ for a collection of pre-main-sequence stars with measured
magnetic field strengths \cite{jkrull}. The annulus where cosmic rays
are accelerated through reconnection and flaring activity is expected
to extend mostly inward from $\rmag$ \cite{lee1998}; to illustrate the
extent of the enrichment region, the vertical cyan lines show
locations corresponding to $\rmag/2$ for the same sample.

Note that the present-day locations of the planets (as shown in the
figure) are not necessarily where they were formed. If rocky planets
form at larger radial distances from their host stars, and
subsequently migrate inward, then they would not be enriched in carbon
through the mechanism considered here.

The stellar hosts for the planets shown in Figure \ref{fig:rmag} have
masses in the range $M_\ast\approx$ $0.10-1.35M_\odot$, whereas the
stellar masses for the $\rmag$ sample have $M_\ast\approx0.3-2$
$M_\odot$. The stellar masses in both data sets thus span the same
range and include the low mass stars of interest (where the habitable
zones are roughly coincident with the disk truncation locations).
Moreover, neither the planet masses $M_p$ or the truncation radii
$\rmag$ show any obvious trends with stellar mass.

Since we are interested in potentially habitable planets around
smaller stars, it is important to note that M stars are generally more
active than solar-type stars and are subject to intense flaring
activity. Some authors have speculated that such energetic activity
could be detrimental for planetary habitability. Although the
conditions required for life remain unknown, this complication should
be noted (see the discussion of \cite{chen2021} and references
therein). On the other hand, the enhanced flaring activity of 
these small stars could result in enhanced spallation rates and 
somewhat larger carbon yields. 

\section{Conclusion} 
\label{sec:conclude} 

This paper presents a new possible solution to the triple alpha
fine-tuning problem by considering how spallation changes the chemical
composition of rocky planets forming in tight orbits around low-mass
stars. If the fundamental constants of nature had different values, so
that the Hoyle resonance in the carbon nucleus had a different energy
level, then carbon production could be greatly reduced. This paper
shows that even in universes that fail to produce carbon by stellar
nucleosynthesis, some carbon is produced by spallation outside of
stars. This argument can be summarized as follows:

If stars fail to produce substantial carbon yields, the nuclei are
processed instead into oxygen and other alpha elements. These larger
nuclei can be broken down into carbon by cosmic radiation through
spallation reactions, thereby providing an alternate source of 
carbon. 

Cosmic ray acceleration takes place near the magnetic truncation
radius $\rmag$ in the disks associated with young stellar objects.
Rocky planets readily form in this region, which is roughly located at
$a\sim\rmag\sim0.1$ AU (Figure \ref{fig:rmag}).  This region
corresponds to the habitable zone for stellar hosts with masses
$M_\ast\sim0.25M_\odot$, which are the most common stars in our
universe. As a result, the raw materials that make up this class of
habitable planets are naturally irradiated by local (stellar) cosmic
rays and are subject to spallation.

The amount of carbon produced depends on the timing of events that
take place during the process of planet formation. For the expected
cosmic ray luminosities, if the rocky material remains optically thin
for a typical time scale of 1 Myr, then the resulting carbon to oxygen
ratio [C/O] $\sim10^{-3}$, comparable to the value inferred for Earth.
However, this value is only realized for relatively small total masses
$M_s\ll M_\oplus$. Longer exposure times lead to higher carbon yields,
whereas larger masses $M_s$ or larger rock sizes $b$ lead to greater
attenuation of the cosmic ray flux and hence smaller carbon yields
(Figure \ref{fig:coratio}). For example, if the total mass in solids
in the reconnection region is comparable to Earth, then an exposure
time of 10 Myr leads to [C/O] $\approx0.0003$, near the low end of 
the range of inferred values for Earth. 

The results of this work have a number of implications: Because
spallation can produce carbon in universes where the triple alpha
process is inoperative, the fine tuning of the fundamental constants
necessary for a viable universe is less severe than previously
claimed. On the other hand, this spallation mechanism tends to produce
only modest amounts of carbon, [C/O] $\sim10^{-4}-10^{-3}$, values 
roughly consistent with Earth abundances, but much smaller than that
of the Sun or the universe as a whole. In addition, carbon production
takes place within the habitable zones of smaller stars with masses
$M_\ast\sim0.25$ $M_\odot$. Although such hosts are the most common
stars in our universe, the habitable zone for solar-type stars lies
farther away from the cosmic ray source, so that direct analogs of our
solar system would remain carbon poor. At the present time, we do not
know exactly how much carbon is required for a planet, or a universe,
to be habitable. Given the modest levels of carbon produced via
spallation, however, it is likely that such universes would be
somewhat less habitable than our own.

Another lesson from this consideration of spallation is that universes
--- including our own --- have a number of possible channels to reach
habitability. For example, a large number of different astrophysical
sources contribute to the production of carbon and other elements,
including supernova explosions, collisions of compact objects, red
giant winds, spallation, and big bang nucleosynthesis (although the
latter tends to make only light nuclei).  If one channel of carbon
production is unavailable, then astrophysics (often) provides
alternate mechanisms.  In the present context, spallation can produce
carbon in other astrophysical settings, including the interstellar
medium and on the surfaces of mature planets orbiting main sequence
stars (as outlined in \ref{sec:alternate}).

Finally, we note that spallation from locally produced cosmic rays can
have important implications for the chemical enrichment and early
evolution of planets in our universe. In this context, carbon is
already abundant, so any additional carbon produced via spallation is
negligible. Spallation also produces short-lived radioactive nuclei,
such as $^{26}$Al, $^{41}$Ca, $^{53}$Mn, $^{138}$La, $^{10}$Be, and
others. These radionuclides provide important sources of heating and
ionization during planet formation, and meteoritic evidence indicates
that our solar system was enriched in these nuclei relative to cosmic
abundances. A great deal of past work \cite{shu1997,lee1998,desch2010}
has focused on radioactive enrichment mechanisms for solar system
meteorites, at distances of order $\sim$few AU. Since many rocky
planets form much closer to their host stars (e.g., Figure
\ref{fig:rmag}), they are expected to experience much greater levels
of radioactive enrichment, which can influence the formation and 
properties of such systems \cite{asradio}. 

\vskip0.15truein
\noindent
{\bf Acknowledgments:} 
We would like to thank Konstantin Batygin, Juliette Becker, George
Fuller, Evan Grohs, Alex Howe, Jonathan Lunine, Martin Rees, and Chris
Spalding for many useful discussions. We also thank an anonymous
referee for many useful comments that improved the manuscript.  This
work was supported by the Leinweber Center for Theoretical Physics and
by the University of Michigan.

\appendix

\section{Alternate Scenarios for Carbon Production}
\label{sec:alternate} 

As outlined in the main text, intense cosmic ray production is
expected to continue only while circumstellar disks remain intact,
typically for several Myr. After this time, the planets are largely
formed, and the magnetic reconnection region at the inner disk edge is
no longer present.  Over longer spans of time, however, carbon
production can take place on the surfaces of completed planets, or in
the interstellar medium. Spallation in the interstellar medium leads
to [C/O] $\sim10^{-7}$ over a Hubble time \cite{adams2019}. This
estimate assumes present-day cosmic ray fluxes, although they could be
somewhat larger in the past (since cosmic rays are produced through
supernovae, which trace the star formation rate, which was larger in
the past). The flux could also be larger in other galaxies in other
universes.  In any case, the interstellar medium levels of [C/O] are
much smaller than the abundances realized in the reconnection regions
of circumstellar disks. Continued irradiation of mature planets leads
to similarly low [C/O] levels, as considered in this section.

After disk dissipation, when planets are fully formed, the host stars
continue to generate cosmic radiation near their stellar surfaces,
with a reduced particle luminosity. In this context, however, we are
interested in small stars, M dwarfs with $M_\ast\sim0.25M_\odot$,
which are observed to have more energetic flaring activity than the
Sun. As a result, the expected cosmic ray fluxes from M dwarfs will be
significantly larger than solar values during the main sequence phase
\cite{vakili,griessmeier} (but smaller than the fluxes generated at
$\rmag$ during the pre-main-sequence phase). More specifically,
estimates for the cosmic ray flux from M dwarfs indicate that the
cosmic ray flux $\Phi\sim10^5$ cm$^{-2}$ s$^{-1}$, for particles with
energy $E>E_0=10$ MeV and for a distance $r$ = 0.1 AU from the star
\cite{griessmeier}.  This estimate assumes that the M dwarf
continually produces energetic flares (which lead to cosmic ray
acceleration) and thus corresponds to the upper end of the possible
range.  With this flux, the coefficient for the carbon production rate
in equation (\ref{rockrate}) becomes $\gamma_0\sim300$ atoms
cm$^{-3}$ s$^{-1}$.

Now consider a planet orbiting a small star at 0.1 AU and assume that
it is exposed to cosmic rays over time $\texp$. The planet is likely
to be spinning, but at any given time an area $A\approx\pi R_p^2$
of the planetary surface will be exposed to the flux of energetic
particles.  The total number of carbon nuclei produced through
spallation takes the approximate form 
\be
N_{\rm C12} \approx (\pi R_p^2) \zeta \gamma_0 \texp 
\sim 10^{38} \left({R_p\over R_\oplus}\right)^2  
\left({\texp\over10\,{\rm Gyr}}\right) \,. 
\ee
For comparison, the biosphere of Earth contains 400 -- 800 billion
tons of carbon, which is equivalent to about 2 -- 4 $\times10^{40}$
carbon atoms. As a result, the number of carbon nuclei that can be
produced by spallation over a Hubble time corresponds to less than 
one percent of the carbon content of our biosphere. With the assumed
cosmic ray flux, the star emits a total number of cosmic rays
$N_{CR}\sim10^{48}$ and the planet intercepts $N_p\sim4\times10^{40}$
over the time $\texp$, so that $N_{CR}\gg N_p\gg N_{\rm C12}$.

The earlier phase of stellar evolution can produce carbon to oxygen
ratios up to [C/O] $\approx0.00025$ for $M_s=1M_\oplus$ (see Figure
\ref{fig:coratio}). Assuming an initial oxygen mass fraction $\xoxy$ =
0.3, the total carbon content for an Earth-like planet corresponds to
$N\sim2\times10^{46}$ nuclei. This value far exceeds that total amount
of carbon produced through the long term channel of spallation. On the
other hand, long term exposure produces carbon at the planetary
surface, where it could be more biologically useful.

Looking at these results another way, the number of carbon nuclei
produced through long-term enrichment corresponds to the number
produced through early enrichment in the upper $\sim1$ cm of the
planet (in the absence of a planetary magnetic field strong enough to
shield the planetary surface from cosmic rays).  This thickness is
comparable to the scale height for spallation (assuming $\zeta\sim1$
cm). As a result, long-term irradiation roughly doubles the carbon
content of the upper layer of the planet (1 cm thick).


\end{document}